\documentclass[12pt]{article}
\usepackage[T1]{fontenc}
\usepackage[utf8]{inputenc}
\usepackage{lmodern}
\usepackage[margin=1in]{geometry}
\usepackage{amsmath,amssymb,amsthm,mathtools}
\usepackage{booktabs,array,tabularx}
\usepackage{enumitem,microtype,xcolor,needspace}
\usepackage[authoryear,round]{natbib}
\usepackage[colorlinks=true,linkcolor=black,citecolor=black,urlcolor=black]{hyperref}
\usepackage[nameinlink,noabbrev]{cleveref}
\crefname{appendix}{appendix}{appendices}
\Crefname{appendix}{Appendix}{Appendices}
\hypersetup{pdftitle={Moral Inversion: Expressive Payoffs and the Reliability of Consensus},
 pdfauthor={Ronen Gradwohl, Arye Hillman, Rok Spruk},
 pdfsubject={Expressive incentives, public information, and incorrect consensus}}
\numberwithin{equation}{section}
\newtheorem{theorem}{Theorem}
\newtheorem{proposition}{Proposition}
\newtheorem{lemma}{Lemma}
\newtheorem{corollary}{Corollary}
\theoremstyle{definition}
\newtheorem{definition}{Definition}
\newtheorem{assumption}{Assumption}
\theoremstyle{remark}

\newcommand{\PP}{\mathbb P}
\newcommand{\EE}{\mathbb E}

\newcommand{\ind}{\mathbf 1}

\newcommand{\Law}{\operatorname{Law}}

\setlist[enumerate]{itemsep=3pt,topsep=5pt}
\title{Moral Inversion:\\
Expressive Payoffs and the Reliability of Consensus}
\author{%
Ronen Gradwohl\thanks{%
Department of Economics, University of Haifa,
199 Aba Khoushy Avenue, Haifa 3498838, Israel.
Email: \href{mailto:rgradwohl@econ.haifa.ac.il}
{rgradwohl@econ.haifa.ac.il}.
Corresponding author.
Financial support from the Israel Science Foundation
(grant number 2039/24) is gratefully acknowledged.}
\qquad
Arye Hillman\thanks{%
Department of Economics, Bar-Ilan University,
Ramat Gan 5290002, Israel.
Email: \href{mailto:arye.hillman@biu.ac.il}
{arye.hillman@biu.ac.il}.}
\qquad
Rok Spruk\thanks{%
School of Economics and Business, University of Ljubljana,
Kardeljeva plo\v{s}\v{c}ad 17, 1000 Ljubljana, Slovenia.
Email: \href{mailto:rok.spruk@ef.uni-lj.si}
{rok.spruk@ef.uni-lj.si}.}
}
\date{\today}
\begin{document}
\maketitle
\begin{abstract}
Moral inversion is a public consensus on an account that falsely
reverses responsibility for a harm. We study how expressive
payoffs affect its emergence in a sequential social-learning
model. A larger payoff makes an account more attractive to
endorse, but also makes each endorsement of that account weaker evidence
in its favor.  We show that under a single-peakedness condition on private
beliefs a larger payoff both increases the probability of an
incorrect consensus on the favored account and reduces its
reliability. Without this condition, both comparisons can
reverse, and the informativeness of the public record can strictly improve in the Blackwell order. Finally, we show that a sufficiently large increase in the expressive payoff nevertheless
restores the benchmark comparative statics. 
\end{abstract}

\section{Introduction}\label{sec:intro}
Moral inversion occurs when those responsible for a harm receive approval while those who suffered the harm are blamed. For example, during the Dreyfus affair, the innocent Alfred Dreyfus was condemned as a traitor while the actual spy, Ferdinand Esterhazy, was acquitted and publicly applauded \citep{Bredin1986,FrenchJustice2011}. Similarly, perpetrators of racial terror lynchings in the United States could receive public acclaim while Black victims were
blamed for the violence inflicted on them \citep{EJI2017}. 
How can such an incorrect attribution gain widespread acceptance, even when contrary information is available?

One explanation is that individuals obtain an expressive, reputational, or professional payoff from endorsing a particular account. They may then endorse the account even when they regard it as less likely to be correct. At a fixed belief, increasing this payoff makes the account's endorsement more attractive. It is then natural to expect that it also makes an incorrect consensus in public endorsement more likely.

However, this natural prediction overlooks a countervailing force, namely,
the information conveyed by an endorsement. When the incentive to endorse
increases due to an expressive payoff, the strength of the information in
favor of the account conveyed by the endorsement decreases. An expressive
payoff to a particular endorsement thus affects both the behavior and the
information it contains. The two effects operate through different channels:
The incentive effect expands the region of public beliefs supporting a public consensus
on the favored account and contracts the region supporting the other consensus.
The informativeness effect changes the size of individuals' belief updates, and so how
far beyond the boundaries of those regions their beliefs land when learning stops.

We study these competing effects in a simple sequential social-learning model. The state is either $H$ or $L$. Each individual observes previous endorsements and receives private information about the state, and then chooses which state to endorse. Individuals value endorsing the correct state, but receive an additional, commonly known expressive payoff for endorsing $L$. A standard result in this model is that individuals eventually all
endorse the same account, and that public beliefs converge to a cascade region in which endorsements are independent of private
information \citep[see, e.g.,][]{SS2000}. We think of moral inversion as an eventual consensus on $L$---that is, all individuals endorsing $L$---when the realized state is $H$. We focus on two outcomes: (i) the probability of an incorrect consensus, $\PP(\text{consensus is }L ~|~ \text{state is }H)$, and (ii) the reliability of the consensus, namely, the probability that the $L$ consensus is correct, $\PP(\text{state is }L ~|~ \text{consensus is }L)$.

We begin with a benchmark in which, once learning has begun, beliefs cannot
enter a cascade region at a finite date. Following the literature, we call
this property \emph{no delayed cascades} \citep[see, again,][]{SS2000}.
Under this assumption, terminal public beliefs equal the cascade boundaries,
so only the incentive effect matters for eventual outcomes. The informativeness
of individual endorsements affects how beliefs approach the boundaries, but not
where they end up. We show that a larger expressive payoff then both raises
the probability of an incorrect $L$ consensus and lowers the reliability of an $L$
consensus.

We then ask, when does the no-delayed-cascades assumption hold? We provide a sufficient condition on the individuals' private information that implies it: roughly, that strong evidence is less likely than weak evidence. 
More specifically, starting from prior \(1/2\), the density of the posterior induced by the private signal weakly decreases with distance from it. This single-peakedness condition is easy to interpret, and includes uniform, triangular, and symmetric truncated normal densities.\footnote{Other incomparable conditions have appeared in the literature, and we discuss them in \Cref{app:otherconditions}.}

Despite its natural interpretation, the condition can fail, and beliefs can
then jump past a cascade boundary. The informativeness effect then matters for
our outcomes of interest, because it determines how far beyond the boundary beliefs land. To drive home the point, our
next result is a simple example in which increasing the expressive payoff has precisely the opposite effect on outcomes than expected: the probability of an incorrect $L$
consensus \emph{decreases}, whereas its reliability \emph{increases}. Moreover,
we show that, in this example, a larger expressive payoff strictly improves the
information conveyed by the history of endorsements in the Blackwell order. In
the example, the higher payoff makes an initial $L$ endorsement weak enough
that it no longer triggers a cascade on its own. An $L$ consensus then forms
only after additional evidence for $L$, and rests, on average, on beliefs
further beyond the boundary. These comparative statics demonstrate that the informativeness
effect, operating through overshooting the cascade boundaries, can strictly dominate the incentive effect.

Our final result is that, even when our single-peakedness condition fails, overshooting the cascade boundaries can
offset the incentive effect only to a limited extent. That is, for each
information structure there is a finite bound on how far terminal beliefs can
overshoot. Thus, if the
increase in the expressive payoff shifts the boundaries by more than this bound,
then the incentive effect once again
dominates the informativeness effect. In this case, the probability of an incorrect consensus
increases, and its reliability decreases. This also implies that the same
information structure can produce opposite comparative statics for smaller and
larger expressive payoff increases.

Expressive incentives thus affect consensus through two competing forces. A
larger expressive payoff makes the favored account more attractive to endorse, but it also
makes an endorsement of that account weaker evidence in its favor. The incentive
effect expands the $L$-cascade region and contracts the $H$-cascade region, while
the informativeness effect determines how far beyond the boundaries beliefs land. Which effect prevails
depends on the structure of private information (in particular, on whether
and how far beliefs can overshoot cascade boundaries) and on the size of the payoff change. Stronger expressive
incentives can therefore reinforce moral inversion, but can also make an
incorrect attribution less likely to become the consensus. Understanding moral
inversion thus requires attention to both the incentives behind public
consensus, and the private evidence that consensus conceals.

We note that our model abstracts from much that characterized historical episodes of
moral inversion, including coercion, prejudice, and the suppression of
evidence. We study a single feature, namely, that endorsing the account that shifts responsibility from perpetrator to victim carries an expressive or
reputational reward. We also assume that observers understand this reward
and discount endorsements accordingly. Our results therefore show that moral
inversion does not require gullible observers. It can emerge, and
expressive rewards can make it more likely, even when everyone interprets
public endorsements correctly.

\paragraph{Related literature}\label{sec:literature}
We use the observational-learning framework of \citet{Banerjee1992}, \citet{BHW1992}, and \citet{SS2000}. Individuals learn from earlier actions, but an action pools several private signals. This leads to herd behavior, in which all individuals take the same action from some point on. Such actions can eventually cease to reveal information. \citet{BHTW2024} survey this literature. We hold the private experiment fixed and vary a common payoff for one action.

A close predecessor is \citet{HSW2012}, who obtain closed-form herd probabilities under a restriction preventing beliefs from jumping into cascades and note that incentive changes alter action informativeness. Our benchmark follows this approach. We characterize a benchmark for monotone comparisons, construct simultaneous reversals in the favored error and consensus reliability with strict Blackwell improvement, and give a quantitative sufficient condition for recovering monotonicity despite overshooting.

\citet{SST2021} connect posterior monotonicity to conditional log-concavity of private log-likelihood ratios and rule out delayed cascades. \citet{HH2012} give related sufficient conditions using hazard and failure ratios. Single-peakedness is another route to the boundary benchmark and need not give monotone updates at every interior belief. A fixed expressive payoff can be absorbed into a shift of normalized initial log odds (see \Cref{sec:comparisons,lem:normalization}), so our state-conditional comparisons also speak to how herd outcomes depend on the prior. The distinction from prior monotonicity is therefore the terminal outcome compared, rather than a new learning dynamic. Our reliability comparisons have no such exact counterpart, because the expressive payoff shifts behavior without shifting the probability of the true state.

Non-monotone effects of information are also established in the
social-learning literature. \citet{PastinePastine2006} provide
numerical examples in which improving the accuracy of asymmetric
binary signals increases the probability of an incorrect cascade.
\citet{BurguetVives2000} show that more precise initial public
information can reduce welfare when private information is costly
to acquire. \citet{Zhang2021} shows that, comparing histories of
equal length, more actions favoring a state can induce a lower
posterior belief in that state. Our comparisons hold the private
experiment fixed and vary a common action payoff, examining both
terminal errors and the informativeness of the resulting public
record.

Our paper also relates to the literature on reputational herding.
\citet{ScharfsteinStein1990} show that managers concerned about
their perceived ability may imitate earlier investment decisions
and disregard private information.
\citet{OttavianiSorensen2001} study sequential debate among
experts whose concern for their reputation can similarly lead
them to suppress private information. \citet{Morris2001} also
studies how reputational concerns can impair communication.
In these models, incentives arise from observers' inferences
about the speaker's type. More broadly, \citet{Hillman1998}
discusses how political correctness shapes economic discourse.
We take a common, known payoff for endorsing a particular
account as given and study how varying this payoff changes
learning. Our focus is on its competing incentive and
informativeness effects, and their implications for the
probability of an incorrect consensus, the reliability of the
favored consensus, and the informativeness of the public record.

Other related applications consider strategic disclosure \citep{BV2025} and informational ambiguity \citep{Chen2026}.
\citet{GPR2006} study how preference heterogeneity can sustain learning, and \citet{BW2001} identify informational benefits of overconfidence. Our examples instead change a common, known incentive among agents with identical preferences. We use the order of \citet{Blackwell1953} to compare the resulting public experiments. \citet{SatoShimizu2026} compare the social value of underlying information structures. In contrast, our private information structure is unchanged, while incentives change what actions reveal.

Our analysis also relates to work distinguishing incentives from
beliefs. \citet{Suen2004} shows how the endogenous choice of
advisers providing coarse information can reinforce prior beliefs.
\citet{EffersonMcKayFehr2020} distinguish the effects of asymmetric
error costs from those of distorted beliefs and examine their
interaction with social learning.

Finally, the application to moral inversion relates to preference falsification \citep{Kuran1995}, conformity \citep{Bernheim1994}, and expressive behavior \citep{Hillman2010}. We focus on how a commonly understood expressive payoff for an endorsement changes the information passed to later individuals and the resulting consensus.

\paragraph{Organization} 
The rest of the paper is organized as follows.  \Cref{sec:model} presents the model. \Cref{sec:benchmark} gives the benchmark and its single-peaked sufficient condition, and \Cref{sec:reversals} presents the example with the opposite comparative statics when the condition fails. \Cref{sec:large} contains the result on comparative statics without the condition, and \Cref{sec:conclusion} concludes.
All proofs and additional technical results are in the appendix.

\section{Model}\label{sec:model}
The state is $\theta\in\{H,L\}$, with common prior $\Pr(\theta=H)=\rho\in(0,1)$. Individuals $t=1,2,\ldots$ act once, in sequence. Individual $t$ observes the action history $\mathbf a^{t-1}=(a_1,\ldots,a_{t-1})$ and a private signal $x_t\in S$, then chooses an endorsement $a_t\in\{H,L\}$. Her utility is
\begin{equation}\label{eq:utility}
u^v(a_t,\theta)=\ind\{a_t=\theta\}+v\ind\{a_t=L\},
\end{equation}
where $0\le v<1$ is an expressive payoff that rewards endorsement of $L$ independently of the state. All individuals have the same preferences, and the prior, payoff, and private experiment are common knowledge.

A  \emph{private signal experiment} consists of a signal space $S$ and two conditional probability laws $F_H,F_L$. Conditional on $\theta$, the signals $(x_t)_{t\ge1}$ are independent draws from $F_\theta$. We assume that $F_H$ and $F_L$ are mutually absolutely continuous: an event has probability zero under one state if and only if it does under the other. The signal likelihood ratio is
\begin{equation}\label{eq:likelihood}
\lambda(x)=\frac{dF_H}{dF_L}(x).
\end{equation}
When the laws have densities, this is their state-$H$ to state-$L$ density ratio, so, for example, a likelihood ratio of three means that the evidence is three times as likely under $H$ as under $L$. We assume bounded evidence \citep{SS2000}, where the likelihood ratio has essential infimum \(1/R\) and essential supremum \(R\), for $R>1$.
These endpoints describe the strongest evidence that signals can approach with positive probability. We generally assume that the distribution of $\lambda$ is atomless under both states, although its support may have gaps.

It is sometimes convenient to describe the private experiment in terms of the beliefs it induces starting with prior $1/2$. To this end,
define the \emph{reference private posterior}
\begin{equation}\label{eq:reference}
p(x)=\frac{\lambda(x)}{1+\lambda(x)}.
\end{equation}
Specifically, this is the posterior belief that signal $x$ would induce from prior $1/2$, without any public history. 
For simplicity, we also assume symmetry, so that
\begin{equation*}\label{eq:symmetry}
\Law(p(x_t)\mid H)=\Law(1-p(x_t)\mid L).
\end{equation*}
By the boundedness assumption, $p$ lies in $[a,1-a]$, where $a=\frac1{1+R}$.

When $p$ has a density under the equal-prior mixture $(F_H+F_L)/2$, we denote it by $h$. In this case we let $f_H,f_L$ denote the conditional densities of the reference posterior. Bayes' rule and symmetry give
\begin{equation}\label{eq:densities}
h(p)=h(1-p),\qquad f_H(p)=2ph(p),\qquad f_L(p)=2(1-p)h(p).
\end{equation}
Conversely, any symmetric probability density $h$ on $[a,1-a]$ defines a private experiment through these formulas. 

\paragraph{Inference from endorsements}
We write $\PP_v$ for the joint law of the state and public record under payoff $v$, and $\PP_{\theta,v}$ for its conditional law given $\theta$. Public beliefs and odds after $t$ actions are
\begin{equation}\label{eq:publicbelief}
\pi_t=\PP_v(\theta=H\mid\mathbf a^t)\qquad\text{and}\qquad O_t=\frac{\pi_t}{1-\pi_t}.
\end{equation}
where $\pi_0=\rho$. We suppress $v$ in $\pi_t$ and $O_t$ when $v$ is clear from context. Individual $t$'s posterior odds are $O_{t-1}\lambda(x_t)$, due to the conditional independence of her private experiment. 

Define the threshold odds
\begin{equation}\label{eq:threshold}
k(v)=\frac{1+v}{1-v}.
\end{equation}
Given expressive payoff $v$, the optimal endorsement is
\begin{equation}\label{eq:decision}
a_t=H\quad\Longleftrightarrow\quad O_{t-1}\lambda(x_t)\ge k(v).
\end{equation}
We assume that ties are resolved in favor of $H$, but note that they have probability zero in the atomless case. Equivalently, an individual endorses $H$ when her posterior belief is at least $(1+v)/2$. Observe that an endorsement of $L$ can be optimal even when the private belief is such that $H$ is more likely.

For a public odds value $O$, let $I_H^v(O)=\{x:\lambda(x)\ge k(v)/O\}$ and $I_L^v(O)=S\setminus I_H^v(O)$ be the two pools of signals that induce actions $H$ and $L$, respectively. The public odds following a positive-probability action $j$ are then
\begin{equation}\label{eq:publicupdate}
T_j^v(O)=O\frac{F_H(I_j^v(O))}{F_L(I_j^v(O))},\qquad j\in\{L,H\}.
\end{equation}
This is the updating rule used throughout the paper. A larger payoff raises the cutoff for endorsement $H$ and makes endorsement $L$ more likely at a fixed belief. It also brings more favorable evidence into the $L$ pool. An $L$ endorsement then provides weaker evidence against $H$, while an $H$ endorsement provides stronger evidence for $H$. In 
\Cref{app:setup} we write out each action likelihood ratio as an average of private likelihood ratios, making these two informational effects explicit.\footnote{See Equations~\eqref{eq:updateH}--\eqref{eq:updateL}.}

\subsection{Herds and outcomes}\label{sec:comparisons}
Observe that when $O\le k/R$, action $L$ is chosen almost surely regardless of the individual's signal, because even the strongest evidence will not suffice to induce odds greater than $k$. Analogously, when $O\ge kR$, action $H$ is chosen almost surely. These are \emph{cascade regions}, since the chosen action does not convey additional private information. The cascade regions' respective boundary beliefs are
\begin{equation}\label{eq:boundaries}
b(v)=\frac{k}{R+k},\qquad d(v)=\frac{kR}{1+kR}.
\end{equation}
Note that both rise with $v$: A larger expressive payoff allows an $L$ cascade with less evidence for $L$, whereas an $H$ cascade requires more evidence for $H$. Regardless of $v$, since public beliefs are a bounded martingale they converge to a terminal belief, which we denote by $X_v=\lim_t\pi_t$. 

For $j\in\{L,H\}$, define the \emph{eventual herd} event
\[
\mathcal J_v=\{\exists T<\infty\text{ such that }a_t=j\text{ for every }t\ge T\},
\]
writing $\mathcal L_v$ and $\mathcal H_v$ for the respective directions.  
We restrict comparisons to payoffs at which the prior is initially informative:
\begin{equation}\label{eq:active}
b(v)<\rho<d(v),\qquad\text{or, equivalently,}\qquad k/R<\rho/(1-\rho)<kR.
\end{equation}
Under this restriction both herds then have positive probability under either state. At prior $1/2$, the restriction is $1\le k<R$. \Cref{app:convergence} proves these standard convergence and conditioning facts.

The main outcomes in which we are interested are
\begin{equation}
e_H(v)=\PP_v(\mathcal L_v\mid \theta=H)\qquad\text{and}\qquad
C_L(v)=\PP_v(\theta=L\mid\mathcal L_v)=1-\EE_v[X_v\mid\mathcal L_v].
\end{equation}
The first is the probability that the favored account becomes the consensus (i.e., there is a herd on $L$) when it is false (i.e., when the state is $H$). The second is the reliability of an $L$ consensus---the probability that the state is actually $L$ given a herd on action $L$. 
We also use
\[
e_L(v)=\PP_v(\mathcal H_v\mid \theta=L)\qquad\text{and}\qquad
e(v)=\rho e_H(v)+(1-\rho)e_L(v)
\]
for the opposite error and aggregate error, where the latter is a measure of accuracy.

The \emph{public-history experiment} is the pair of conditional laws of the record $(a_1,a_2,\ldots)$ under $H$ and $L$. The \emph{herd experiment} is the same, but restricted to only the eventual herd direction. An experiment Blackwell dominates another if a state-independent transformation of its observation, possibly randomized, reproduces the other's conditional laws under both states. Strict dominance means that no such transformation works in reverse. 

\paragraph{Expressive payoffs as a prior shift}
We note that a common, known expressive payoff acts on behavior exactly like a
shift in the prior. Individual $t$ endorses $H$ if and only if
$(O_{t-1}/k(v))\,\lambda(x_t)\ge 1$, so the action sequence under
prior $\rho$ and payoff $v$ has the same state-conditional law as in
a standard model with no expressive payoff and prior odds
$\rho/[(1-\rho)k(v)]$. The two error probabilities, the herd
experiment, and the public-history experiment are therefore those of
a standard herding model with a prior tilted toward $L$, and our
results on these objects can equally be read as comparative statics
in the prior. However, this is not the case for reliability. The expressive payoff tilts
the prior that governs behavior, but not the probability of the true
state against which the consensus is judged. By Bayes' rule at the
true prior,
\[
  \frac{C_L(v)}{1-C_L(v)}=\frac{1-\rho}{\rho}\cdot
  \frac{1-e_L(v)}{e_H(v)},
\]
so at a fixed $\rho$, $C_L$ changes with $v$ only insofar as an $L$
herd becomes stronger or weaker evidence for $L$. If instead there were a shift of
the prior toward $L$, this would in addition mechanically raise the reliability of an
$L$ herd. The contrast is clearest in our monotone benchmark (in \Cref{sec:benchmark}): the reliability of an $L$ herd in the standard model is
$1-b(0)$ at every initially informative prior, but in our benchmark setting the reliability
$C_L(v)=1-b(v)$ strictly falls with $v$.

\section{The monotone benchmark}\label{sec:benchmark}
In this section we analyze the effect of $v$ on the probability of a false consensus and on the reliability of that consensus under the oft-used assumption of \emph{no-delayed-cascades} on the learning dynamics. We show that, under this assumption, the comparative statics go in the ``expected'' direction.
\subsection{No delayed cascades}
A delayed cascade is one that begins after learning has started \citep[see, e.g.,][Corollary~4]{SST2021}. 
\begin{definition}[No delayed cascades]\label[definition]{def:NDC}
The private experiment has no delayed cascades if, for every $v\in[0,1)$,
\begin{equation}\label{eq:NDC}
\frac{k(v)}R<T_j^v(O)<k(v)R
\quad\text{for every }O\in\left(\frac{k(v)}R,k(v)R\right),\quad j\in\{L,H\}.
\end{equation}
\end{definition}
Under no-delayed-cascades, both actions leave an informative belief informative. We could replace the strict inequalities by weak inequalities, leaving the benchmark comparisons unchanged. We refer to the resulting property as \emph{no overshooting}.
We note that this requirement is actually independent of $v$, since dividing odds and updates by $k(v)$ gives a common learning region and updating rule (this is formalized as \Cref{lem:normalization} in the appendix).

We now state our first result.
\begin{theorem}
\label{thm:benchmark}
Fix a prior $\rho$ and suppose the private experiment has no delayed cascades.
For any $v_1<v_2$ satisfying
\eqref{eq:active},
\[
e_H(v_2)>e_H(v_1)
\qquad\text{and}\qquad
C_L(v_2)<C_L(v_1).
\]
\end{theorem}

The argument is short. Beliefs converge, cannot converge to an informative interior belief, and cannot cross a boundary (because of no-delayed-cascades). The terminal belief must therefore equal the boundary corresponding to the eventual herd, namely, $b(v)$ or $d(v)$. 

Slightly more formally, Bayes' rule  gives
\[
C_L(v)=1-b(v)\qquad\text{and}\qquad
e_H(v)=
\frac{b(v)\bigl(d(v)-\rho\bigr)}
     {\rho\bigl(d(v)-b(v)\bigr)}.
\]
The first expression decreases with $b(v)$, while the second
increases with either boundary whenever $b(v)<\rho<d(v)$.
Since both boundaries increase with $v$, the stated
comparisons follow.

For example, suppose the prior is $1/2$, $R=3$, and there
are no delayed cascades. Increasing $v$ from $0$ to $1/4$
raises $e_H$ from $1/4$ to $1/2$ and reduces $C_L$ from
$3/4$ to $9/14$. Thus the incorrect consensus
becomes more likely, while a consensus on that
account becomes less reliable.

\subsection{Single-peaked private beliefs}\label{sec:primitives}
When does no-delayed-cascades hold? In this section we provide an intuitive sufficient condition on the private experiment that guarantees it. Roughly, our condition says that weak private evidence is relatively common, while stronger evidence is progressively less common.

We note that other, incomparable conditions that also restrict cascade entry have appeared in the literature, such as the log-concavity of the
state-conditional density of private log-likelihood ratios \citep{SST2021}, bounds on the private-posterior density \citep{HSW2012}, and conditions on hazard and failure ratios \citep{HH2012}. We compare them to our condition in \Cref{app:otherconditions}.

Our primitive sufficient condition places relatively more density on less informative private assessments.
\begin{assumption}[Single-peakedness]\label[assumption]{ass:SP}
The reference private posterior $p$ has a density $h$ with essential support endpoints $a$ and $1-a$, and $h$ has a version that is nondecreasing on $[a,1/2]$.
\end{assumption}
By symmetry, the density is nonincreasing above $1/2$. That is, the private experiment is such that, when starting from prior $1/2$, the density of the posterior belief weakly decreases with distance from the prior.

Examples of single-peaked experiments include the uniform density on $[a,1-a]$, a triangular density peaked at $1/2$, and symmetric truncated normal densities centered at $1/2$. There is also a direct diagnostic-test interpretation: Suppose an individual observes a binary recommendation and its accuracy $Q\ge1/2$, where the recommendation equals the true state with probability $Q$ and $Q$ is independent of the state. Her reference posterior is $Q$ after an $H$ recommendation and $1-Q$ after an $L$ recommendation. 
The density of $Q$ is $2h(q)$.  \Cref{ass:SP} is then
equivalent to test accuracy having a nonincreasing density, meaning that
weak tests are relatively common.

\begin{proposition}\label[proposition]{prop:primitives}
Under \Cref{ass:SP}, the private experiment has no delayed cascades. Consequently, the conclusions of \Cref{thm:benchmark} apply.
\end{proposition}

To see the intuition, suppose that an individual chooses $H$
when her reference posterior exceeds a cutoff $c\in(a,1-a)$.
An $H$ endorsement reveals only that the private posterior
exceeds $c$, so its informativeness depends on the mean
reference posterior $\mu(c)$ in this pool. If the selected posteriors
were uniform on $[c,1-a]$, their mean would be the midpoint
$(c+1-a)/2$. Symmetry and single-peakedness ensure that their
actual mean is no larger. As public belief approaches the
$H$-cascade boundary, the cutoff falls and the endorsement
pools in weaker evidence. More specifically, the cutoff $c$ corresponds to current public
odds $O=k(1-c)/c$, where $k=k(v)$. The midpoint bound
limits the updated public odds after an $H$ endorsement, $O_H^+$, as follows:
\[
\frac{O_H^+}{k}
=\frac{1-c}{c}\frac{\mu(c)}{1-\mu(c)}
\le \frac{(1-c)(c+1-a)}{c(1+a-c)}
<R.
\]
The final inequality uses $a<c<1-a$ and $R=(1-a)/a$.
Thus, the updated odds remain below the $H$-cascade boundary
$kR$. Symmetry gives the corresponding conclusion after an
$L$ endorsement,
and \Cref{app:primitives} proves both the tail bound and the proposition.

\section{Reversals without no-delayed-cascades}\label{sec:reversals}
What happens if no-delayed-cascades does not hold? In this section we describe an example that shows that, in this case, an increase in the expressive payoff can both reduce the probability of an incorrect $L$ herd and increase the reliability of an $L$ consensus. Both conclusions of \Cref{thm:benchmark} thus fail. Moreover, the example demonstrates that a larger expressive payoff can actually  improve the entire public experiment in the Blackwell order.

Let the prior be $1/2$. Each individual observes a binary test recommendation and its accuracy. The test is $75\%$ accurate with probability $2/7$ and $90\%$ accurate with probability $5/7$, independently of the state. Conditional on accuracy $q$, the recommendation is correct with probability $q$. Tests are independent across individuals, conditional on the state. This gives the following symmetric four-signal experiment: 
\begin{table}[ht]
\centering
\label{tab:signals}
\begin{tabular}{lccc}\toprule
Signal & $\PP(s\mid H)$ & $\PP(s\mid L)$ & Likelihood ratio\\\midrule
Strong $L$ & $1/14$ & $9/14$ & $1/9$\\
Moderate $L$ & $1/14$ & $3/14$ & $1/3$\\
Moderate $H$ & $3/14$ & $1/14$ & $3$\\
Strong $H$ & $9/14$ & $1/14$ & $9$\\\bottomrule
\end{tabular}
\caption{A symmetric four-signal experiment}
\end{table}

Individuals choose $H$ when public odds $O$, multiplied by the private likelihood ratio, are at least $k(v)=(1+v)/(1-v)$. We compare two regimes,
\[
v_1=\frac37,\quad k_1=\frac52,
\qquad\text{and}\qquad
v_2=\frac7{13},\quad k_2=\frac{10}3.
\]
There are no ties on the equilibrium paths in either regime.
Under the lower expressive payoff $v_1$, an initial $L$ endorsement pools the two negative signals. Its likelihood ratio is $1/6$, so public odds fall to $1/6$ and an $L$ cascade begins. Under the higher payoff $v_2$, the moderate $H$ signal also induces $L$. The same endorsement now conveys likelihood ratio $5/13$ and leaves learning active, since
\[
\frac5{13}>\frac{k_2}{9}=\frac{10}{27}.
\]
An initial $L$ endorsement is more likely, but it no longer settles the matter. For example, two consecutive $L$ endorsements lead to odds $25/169$, below the lower-payoff $L$-herd odds $1/6$. The additional endorsement can therefore make the eventual consensus rest on stronger evidence for $L$.

\Needspace{7\baselineskip}
This observation alone does not establish a comparison of eventual outcomes, because the remaining histories also change. The calculation in \Cref{app:finite} gives the following values, rounded to five decimal places:
\begin{equation}\label{eq:outcomes}
\begin{array}{c|cc}
&v_1=3/7&v_2=7/13\\\hline
 e_H(v)&.16279&.15627\\[2pt]
 C_L(v)&.85714&.86296.
\end{array}
\end{equation}

\bigskip
\Needspace{10\baselineskip}
The following proposition formalizes these results.
\begin{proposition}\label[proposition]{prop:confidence}
For the experiment in \Cref{tab:signals}, increasing the expressive payoff from $3/7$ to $7/13$ strictly raises $C_L$ and strictly lowers $e_H$. Moreover, the higher-payoff regime strictly Blackwell dominates the lower-payoff regime both as a public-history experiment and as an eventual-herd experiment. Finally, there is a symmetric, bounded, atomless private experiment that generates exactly the same public-history laws as its discrete counterpart.
\end{proposition}

The reversal holds over intervals of payoffs. For the discrete
experiment, every pair
\[
v_1\in(1/3,1/2),
\qquad
v_2\in(27/53,16/29)
\]
generates the same respective public-history laws as the two
regimes above. Thus the error probabilities, consensus
reliabilities, and strict Blackwell comparisons remain unchanged
throughout these intervals. Appendix~\ref{app:reversalintervals}
verifies this robustness.

The example illustrates two competing effects of expressive incentives. A higher expressive payoff makes individuals more willing to endorse \(L\), but it also changes what others learn from that endorsement. Under the lower expressive payoff, an initial \(L\) endorsement reveals a negative signal and immediately starts an \(L\) cascade. Under the higher expressive payoff, it can also reflect a moderately positive signal, so it provides weaker evidence for \(L\) and leaves subsequent individuals responsive to their private information. An \(L\) consensus can therefore require additional negative evidence. In this example, the resulting improvement in learning outweighs the greater willingness to endorse \(L\): an incorrect \(L\) consensus becomes less likely, and an eventual \(L\) consensus becomes more reliable. The higher expressive payoff thus improves information by weakening the evidence conveyed by an individual endorsement, so that an $L$ consensus rests, on average, on beliefs further beyond the cascade boundary.

The proof of \Cref{prop:confidence} appears in \Cref{app:reversals}. The proof shows that the opposite error also falls, from $e_L(v_1)=1/43$ to $e_L(v_2)=525/32852$. The Blackwell comparison follows by showing that a state-independent randomization of the higher-payoff herd direction reproduces the lower-payoff herd direction. In the lower-payoff regime, terminal public beliefs are constant conditional on each herd, so the rest of its record can then be simulated without knowing the state. This establishes dominance of entire public records. The comparison is strict because the higher-payoff record can support beliefs in $H$ exceeding the lower-payoff record's maximum.

The example reverses both main predictions of the monotone benchmark: the favored incorrect consensus becomes less frequent, while a consensus on the favored account becomes more reliable. Thus, these reversals can coexist with a strict improvement in public information. In \Cref{app:atomless} we then replace the four signals  by narrow intervals of private beliefs, leaving the induced records exactly unchanged, thereby showing that the reversals of \Cref{prop:confidence} can also hold when the likelihood distribution is atomless.

\section{Monotonicity without no-delayed-cascades}\label{sec:large}
The example in \Cref{sec:reversals} shows that, when no-delayed-cascades does not hold, a small change in the expressive payoff can have a counterintuitive effect. In this section we show that a sufficiently large change nevertheless gives a monotone comparison for every symmetric atomless bounded experiment, provided both regimes are initially informative. 
The main idea is to show that overshooting can occur only by a bounded factor determined by the private experiment alone.

For a likelihood-ratio cutoff $c<R$, define $m(c)=\EE_L[\lambda\mid\lambda\ge c]$, and let $m(R)=R$. The expectation here is over the private signal law in state $L$. The {\em overshoot factor} is
\begin{equation}\label{eq:kappadef}
\kappa=\frac1R\max_{c\in[1/R,R]}\frac{m(c)}c.
\end{equation}
If the cutoff is $c$, the current public odds are $k/c$. An $H$ endorsement multiplies these odds by $m(c)$, so its updated odds relative to the $H$-cascade boundary $kR$ are $m(c)/(Rc)$.

The factor $\kappa$ measures how far learning can overshoot a cascade
boundary. Note that $\kappa$ depends only on the private experiment, not on $v$ or the prior. Symmetry gives the corresponding lower bound after $L$. For every
symmetric atomless bounded experiment,
\begin{equation}\label{eq:kapparange}
1\le\kappa<R.
\end{equation}
To interpret $\kappa$, recall that the cascade boundaries in odds
are $k/R$ and $kR$, where $k=k(v)$. Under no delayed cascades, or the weaker no-overshooting condition, terminal
odds equal the boundary corresponding to the eventual herd. Allowing
for overshooting, they can rise above
the $H$ boundary or fall below the $L$ boundary, but by at most a multiplicative factor of $\kappa$:
\begin{equation}\label{eq:bandsbody}
\frac{k}{\kappa R}\le O_\infty\le\frac{k}{R}
\quad\text{on }\mathcal L_v,
\qquad
kR\le O_\infty\le k\kappa R
\quad\text{on }\mathcal H_v.
\end{equation}
When $\kappa=1$, these intervals collapse to the two boundaries,
recovering the no-overshooting benchmark. A larger $\kappa$ allows
terminal beliefs to lie farther beyond those boundaries. The bounds
and the restriction $\kappa<R$ are established in
\Cref{lem:overshoot} in the appendix.

We now state our main result for this section.
\begin{theorem}\label{thm:separation}
Fix the prior and a symmetric atomless bounded private experiment. Let $v_1<v_2$, and suppose  learning is initially possible in both regimes. If
\begin{equation}\label{eq:separation}
\frac{k(v_2)}{k(v_1)}\ge\kappa,
\end{equation}
then
\begin{equation}\label{eq:largeoutcomes}
e_H(v_2)\ge e_H(v_1)\qquad\text{and}\qquad C_L(v_2)\le C_L(v_1).
\end{equation}
Both inequalities are strict if $k(v_2)/k(v_1)>\kappa$.
\end{theorem}

The argument compares the ranges in \eqref{eq:bandsbody}. For an $L$ herd, the smallest higher-payoff odds are $k_2/(\kappa R)$ and the largest lower-payoff odds are $k_1/R$. Condition \eqref{eq:separation} orders them, as well as the $H$-cascade ranges. Both cascade-conditioned mean beliefs therefore rise. The martingale mean and Bayes' rule translate these two increases into the error comparisons.

A larger expressive payoff moves both boundaries toward $H$. Overshooting can offset this movement, but only by a factor of $\kappa$. When the boundary movement dominates that bound, the favored incorrect consensus becomes more likely and an $L$ consensus becomes less reliable. 

\begin{corollary}\label[corollary]{cor:zero}
At prior $1/2$, every symmetric atomless bounded private experiment has a nonempty interval
\begin{equation}\label{eq:zerorange}
\frac{\kappa-1}{\kappa+1}<v<\frac{R-1}{R+1}
\end{equation}
on which $e_H(v)>e_H(0)$ and $C_L(v)<C_L(0)$.
\end{corollary}
The interval is nonempty because $\kappa<R$, and its upper endpoint ensures that private learning is initially still possible. 

Let us return to the four-signal example from \Cref{sec:reversals}. Here $R=9$ and $\kappa=13/5$.
We showed that raising the expressive payoff from $3/7$ to $7/13$ reduces
$e_H$ and raises $C_L$. However, a sufficiently large increase from the
same starting point reverses these comparisons. For example,
take $v_3=3/4$. Since
\[
\frac{k(v_3)}{k(3/7)}
=\frac{7}{5/2}
=\frac{14}{5}>\kappa,
\]
the odds-bound argument underlying \Cref{thm:separation} gives
\[
e_H(3/4)>e_H(3/7)
\qquad\text{and}\qquad
C_L(3/4)<C_L(3/7).
\]
Thus, in the same experiment, a smaller payoff increase decreases the probability of an incorrect consensus and improves
its reliability, whereas a larger
increase recovers the more intuitive ordering from Theorem~\ref{thm:benchmark}.

\section{Conclusion}\label{sec:conclusion}
Expressive payoffs affect both the willingness to endorse an
account and the evidence that others infer from its endorsement.
Under our single-peakedness condition, stronger expressive payoffs make
the favored incorrect consensus more likely and an observed
consensus less reliable. Outside this benchmark, weaker inference
from endorsements can preserve learning and reverse
both comparisons. Sufficiently large increases in the expressive payoff nevertheless
restore the benchmark predictions. Moral inversion can therefore
persist even when individuals understand the incentives behind
public endorsements and update correctly, but its relationship
with those incentives depends on the private evidence available
and the extent to which endorsements continue to reveal it.

\subsection*{Declaration of generative AI and AI-assisted technologies in the manuscript preparation process}

During the preparation of this work, the authors used various versions of ChatGPT and Claude to assist in the analysis and check derivations, to draft sections of the manuscript, and to review the final manuscript for notational consistency. The authors reviewed and edited the output as needed and take full responsibility for the content of the published article. 

\begingroup
\setlength{\bibsep}{4pt}

\endgroup

\clearpage
\appendix
\counterwithin{table}{section}

\begin{center}
    \begin{Large}
        \textbf{Appendix}
    \end{Large}
\end{center}
\section{Learning dynamics and Bayesian accounting}\label[appendix]{app:setup}

\subsection{Updates and normalization}
The signal space is standard Borel. For any positive-probability
signal event $A$, its likelihood ratio is
$F_H(A)/F_L(A)=\EE_L[\lambda\mid A]$. Hence
\begin{align}
O_H^+&=O\EE_L[\lambda\mid\lambda\ge k/O],
\label{eq:updateH}\\
O_L^+&=O\EE_L[\lambda\mid\lambda<k/O],
\label{eq:updateL}
\end{align}
where the former are the updated odds after an $H$ action and the latter after an $L$ action.
Raising the cutoff weakly raises either conditional mean. Thus,
at fixed public odds, the posterior following either given
action weakly increases with $v$.

It will be useful to work with the normalized odds (the odds normalized by $k=k(v)$), for which the updating rules are independent of $v$.
\begin{lemma}\label[lemma]{lem:normalization}
Normalized odds $y=O/k$ follow the updating rules
\begin{equation}\label{eq:normalized}
\Phi_H(y)=y\EE_L[\lambda\mid\lambda\ge1/y],
\qquad
\Phi_L(y)=y\EE_L[\lambda\mid\lambda<1/y].
\end{equation}
These maps are independent of $v$, with initial value
$y_0=\rho/[(1-\rho)k]$. Symmetry and atomlessness imply
\begin{equation}\label{eq:reflection}
\Phi_L(y)=\frac1{\Phi_H(1/y)}.
\end{equation}
\end{lemma}

\begin{proof}
Substitute $O=ky$ into \eqref{eq:updateH}--\eqref{eq:updateL}.
Reflection exchanges the states and replaces $\lambda$ by
$1/\lambda$, giving \eqref{eq:reflection}. Atomlessness makes
strict and weak cutoffs equivalent.
\end{proof}

\subsection{Convergence}\label[appendix]{app:convergence}
The following is a standard result, included here for completeness.
\begin{lemma}\label[lemma]{lem:convergence}
Public beliefs converge almost surely and in $L^1$ to $X_v$,
with $\EE_v X_v=\rho$. Eventual herding occurs almost surely.
Under \eqref{eq:active}, both herd directions have positive
probability under either state, and
\[
X_v\le b(v)\text{ on }\mathcal L_v,
\qquad
X_v\ge d(v)\text{ on }\mathcal H_v.
\]
\end{lemma}

\begin{proof}
Public beliefs form a bounded martingale, giving convergence
and $\EE_v X_v=\rho$. The herding and terminal-belief
conclusions follow from the standard bounded-signal results
of \citet{SS2000}. To apply them here, observe that, conditional
on either state, normalized learning is identical to learning
with zero expressive payoff and prior
\[
\widetilde\rho=\frac{\rho}{\rho+k(1-\rho)}.
\]
Its cascade boundaries in normalized odds are $1/R$ and $R$,
which correspond to actual beliefs $b(v)$ and $d(v)$.

For completeness, both herd directions must have positive
unconditional probability: otherwise the terminal-belief
bounds would contradict $b(v)<\rho<d(v)$ and $\EE_v X_v=\rho$.
Each action multiplies odds by a factor in $[1/R,R]$, so,
under \eqref{eq:active}, odds remain in
$[k/R^2,kR^2]$, including after cascade entry. Thus $X_v$
is bounded away from zero and one. For every event $B$
determined by the entire public record, Bayesian calibration
gives
\begin{equation}\label{eq:calibration}
\PP_v(H,B)=\EE_v[X_v\ind_B],
\qquad
\PP_v(L,B)=\EE_v[(1-X_v)\ind_B].
\end{equation}
Consequently, each positive-probability herd event also has
positive probability under either state.
\end{proof}

\subsection{Conditional means and error probabilities}
\label[appendix]{app:conditioning}

Denote the average public belief conditional on a cascade as $m_j(v)=\EE_v[X_v\mid\mathcal J_v]$, for
$j\in\{L,H\}$.

\begin{lemma}\label[lemma]{lem:accounting}
Under \eqref{eq:active}, $m_L<\rho<m_H$, and
\begin{equation}\label{eq:accounting}
e_H=\frac{m_L(m_H-\rho)}{\rho(m_H-m_L)},
\qquad
e_L=\frac{(1-m_H)(\rho-m_L)}
           {(1-\rho)(m_H-m_L)}.
\end{equation}
The first expression is strictly increasing in each mean,
and the second is strictly decreasing. Also $C_L=1-m_L$.
\end{lemma}

\begin{proof}
Let $w=\PP_v(\mathcal L_v)$. The terminal-belief bounds imply
$m_L\le b(v)<\rho<d(v)\le m_H$. The martingale mean and
\eqref{eq:calibration} give
\[
\rho=wm_L+(1-w)m_H,\qquad
\rho e_H=wm_L,\qquad
(1-\rho)e_L=(1-w)(1-m_H).
\]
Solving yields \eqref{eq:accounting}, and differentiation gives
the stated monotonicity. Finally, calibration gives
$C_L=1-m_L$.
\end{proof}

\section{Proof of the boundary benchmark}\label[appendix]{app:benchmark}
We prove the following, more detailed version of \Cref{thm:benchmark}:
\begin{theorem}\label{thm:benchmark-strong}
Suppose there are no delayed cascades. For every initially informative prior,
\begin{equation}\label{eq:endpoints}
X_v=b(v)\text{ on }\mathcal L_v,\qquad X_v=d(v)\text{ on }\mathcal H_v.
\end{equation}
For a fixed prior and any $v_1<v_2$ satisfying \eqref{eq:active} in both regimes,
\[
e_H(v_2)>e_H(v_1),\qquad C_L(v_2)<C_L(v_1),\qquad e_L(v_2)<e_L(v_1).
\]
At prior $1/2$ and each $1\le k(v)<R$,
\begin{align}
e_H(v)&=\frac{kR-1}{R^2-1},&e_L(v)&=\frac{R/k-1}{R^2-1},\label{eq:benchmarkerrors}\\
C_L(v)&=\frac{R}{R+k},&
e(v)-e(0)&=\frac{2Rv^2}{(R^2-1)(1-v^2)},\label{eq:benchmarkconfidence}
\end{align}
where $e(0)=1/(R+1)$. The same conclusions hold under the weaker no-overshooting condition.
\end{theorem}

\begin{proof}[Proof of \Cref{thm:benchmark-strong}]
Under no delayed cascades, normalized odds remain in $(1/R,R)$ at every finite history. Under the no-overshooting condition they remain in $[1/R,R]$. \Cref{lem:convergence} excludes an interior limit and identifies the limits with the herd directions. This gives \eqref{eq:endpoints} in either case. Both boundary beliefs strictly increase in $v$, so the accounting lemma gives the fixed-prior error comparison and $C_L$ strictly decreases.

At prior $1/2$, substitute $m_L=k/(R+k)$ and $m_H=kR/(1+kR)$ in \eqref{eq:accounting}. This gives \eqref{eq:benchmarkerrors} and $C_L=R/(R+k)$. Averaging the errors yields
\[
e(v)=\frac{R(k+k^{-1})-2}{2(R^2-1)},\qquad
e(v)-e(0)=\frac{R(k-1)^2}{2k(R^2-1)}.
\]
Using $k=(1+v)/(1-v)$ proves \eqref{eq:benchmarkconfidence}. For $1\le k<R$, the terminal beliefs straddle $1/2$, so the herd action also minimizes classification error given the entire record.
\end{proof}

\section{Single-peakedness and comparison with log-concavity}\label[appendix]{app:primitives}
\subsection{Proof of the primitive sufficient condition}
\begin{proof}[Proof of \Cref{prop:primitives}]
Let $h$ satisfy \Cref{ass:SP}. A symmetric single-peaked density is a mixture of uniform densities on intervals centered at $1/2$. To verify this, write
\[
h(p)=\int_0^\infty\ind\{h(p)>z\}\,dz.
\]
Up to null sets, every nonempty superlevel set is an interval centered at $1/2$. Multiplying its normalized uniform density by its length and using Tonelli's theorem gives a mixture with weights integrating to one.

Fix a cutoff $c\in[a,1-a)$. For a component on $[1/2-r,1/2+r]\subseteq[a,1-a]$, if $c$ lies below the component's support, its selected mean is $1/2\le(c+1-a)/2$. If $c$ lies inside its support, its selected mean is $(c+1/2+r)/2\le(c+1-a)/2$. Components wholly below the cutoff have zero selected mass. Averaging over the remaining components, with their conditional selection weights, gives  \begin{equation*}\label{eq:tailmean}
\mu(c)=\frac{\int_c^{1-a}p h(p)\,dp}
             {\int_c^{1-a}h(p)\,dp}
\le\frac{c+1-a}{2}.
\end{equation*}

An $H$ action at normalized public odds $y=O/k\in(1/R,R)$ selects $p\ge c=1/(1+y)$. By \eqref{eq:densities}, the action likelihood ratio is $\mu(c)/(1-\mu(c))$. Hence
\[
\frac{T_H^v(O)}k
=\frac{1-c}{c}\frac{\mu(c)}{1-\mu(c)}
\le\frac{(1-c)(c+1-a)}{c(1+a-c)}<\frac{1-a}{a}=R.
\]
For the last inequality, the difference of the cross-products is
\[
a(1-c)(c+1-a)-(1-a)c(1+a-c)
=(c-a)\bigl[(1-2a)c-(1-a)\bigr]<0
\]
for $a<c<1-a$. All denominators are positive. An $H$ action strictly raises odds at an informative belief, so it also stays above the lower boundary. Reflection \eqref{eq:reflection} gives the corresponding strict lower bound after $L$, which also lowers odds and cannot cross the upper boundary. This proves \eqref{eq:NDC}.
\end{proof}

\subsection{Other conditions in the literature}\label[appendix]{app:otherconditions}
Several signal restrictions lead to the same boundary benchmark through different arguments. We compare them with single-peakedness.

\paragraph{Conditional log-concavity}
The condition (LC) of \citet[][Section~7]{SST2021} requires the state-conditional density $g_\theta(s)$ of $s=\log\lambda$ to be strictly log-concave with interval support. Since $g_H(s)=e^s g_L(s)$, it suffices to check either state. Their condition rules out delayed cascades, and is therefore another route to \Cref{thm:benchmark} in our setting.

The distinction is between shape restrictions on different densities. Our assumption concerns the equal-prior density of private posteriors $p$. Their condition concerns a density of log odds $s$, conditional on the state. Neither implies the other. For example,
\begin{equation}\label{eq:normalexample}
h(p)\propto e^{-64(p-1/2)^2},\qquad p\in[1/16,15/16],
\end{equation}
satisfies \Cref{ass:SP} but violates (LC). Conversely, the U-shaped density $h(p)\propto[p(1-p)]^{-5/4}$ on $[1/8,7/8]$ satisfies (LC) but violates single-peakedness (see \Cref{app:LC}).

\paragraph{A bound on density height}
In the search-goods model of \citet{HSW2012}, Assumption~1 caps the density of a private posterior about whether a purchase is worthwhile. Writing $q_H>q_L$ for the probabilities that the product is worth buying in the two quality states, the cap is
\begin{equation}\label{eq:HSWcap}
f_{\omega_1}(\sigma)\le
\frac{2(1-q_L)(1-q_H)}{q_H-q_L}.
\end{equation}
Their Lemma~1 uses it to prevent beliefs from jumping into the no-search cascade. This is a restriction on density height, with a bound depending on purchasing probabilities. Single-peakedness restricts shape and permits arbitrarily concentrated truncated normals. Moreover, their private evidence concerns purchase value, and consumers observe aggregate purchases rather than our ordered endorsements. The conditions serve a related purpose in different models, but they are incomparable.

Finally, \citet{HH2012} provide related sufficient conditions through hazard and failure ratios. These different approaches underline the role of the private experiment in determining whether learning ends at a boundary or can jump past it. We find our condition, single-peakedness, to be appealing due to its natural interpretation and tractable verification.

\subsubsection{Formal comparison with conditional log-concavity}
\label[appendix]{app:LC}
Let $s=\log[p/(1-p)]$. Its state-$L$ density is
\[
g_L(s)=f_L(p)\frac{dp}{ds}=2p(1-p)^2h(p),
\]
and $g_H(s)=e^s g_L(s)$. Adding the linear term $s$ to a log density does not change strict concavity. Thus the conditional log-concavity condition of \citet{SST2021} is the same under either state.

For a symmetric truncated normal,
\[
h(p)\propto\exp[-(p-1/2)^2/(2\sigma^2)],\qquad p\in[a,1-a],
\]
differentiation gives
\[
\frac{d^2}{ds^2}\log g_L(s)
=p(1-p)\left[-3+\frac{1-6p(1-p)}{2\sigma^2}\right].
\]
The bracketed expression is largest at the support endpoints. Conditional LC therefore holds exactly when $\sigma^2\ge1/6-a(1-a)$, and equality still gives strict concavity because the second derivative is strictly negative in the interior. If the inequality fails, curvature is positive on an interval near an endpoint. In \eqref{eq:normalexample}, $\sigma^2=1/128$ and the curvature at $p=1/8$ equals $133/64>0$. This experiment satisfies single-peakedness and violates LC.

Conversely, let $h(p)\propto[p(1-p)]^{-5/4}$ on $[1/8,7/8]$. It is U-shaped and violates \Cref{ass:SP}, but
\[
g_L(s)\propto p^{-1/4}(1-p)^{3/4},\qquad
\frac{d^2}{ds^2}\log g_L(s)=-\frac12p(1-p)<0,
\]
and so satisfies conditional LC. These two examples show that the restrictions are incomparable.

\section{Reversal example}\label[appendix]{app:reversals}
\subsection{Finite-signal verification}\label[appendix]{app:finite}
\subsubsection{The lower payoff}
Under $k_1=5/2$, the only informative odds are $1$ and $6$, with transitions
\[
1\xrightarrow{L}1/6,\qquad1\xrightarrow{H}6,
\qquad6\xrightarrow{L}1,\qquad6\xrightarrow{H}36.
\]
At both informative states, $H$ pools the two positive signals. Its likelihood ratio is $6$ and its probabilities are $6/7$ under $H$ and $1/7$ under $L$. The states $1/6$ and $36$ absorb strictly: the respective cascade boundaries are $5/18$ and $45/2$.

For either true state, let $b=\PP_\theta(H\text{ action})$ at these two informative states. Let $x$ be the probability of an eventual $L$ herd starting from odds one, and $y$ its probability starting from odds six. Then
\[
x=(1-b)+by,\qquad y=(1-b)x,
\qquad x=\frac{1-b}{1-b(1-b)}.
\]
Substitution of $b=6/7$ or $1/7$ gives
\begin{equation}\label{eq:oldprob}
\PP_H(\mathcal L_{v_1})=\frac7{43},\qquad
\PP_L(\mathcal L_{v_1})=\frac{42}{43}.
\end{equation}
The terminal posteriors are $1/7$ and $36/37$. Hence $C_L(v_1)=6/7$ and $e_L(v_1)=1/43$.

\subsubsection{The higher payoff}
Under $k_2=10/3$, an informative action can pool signals in three ways. In \Cref{tab:types}, the type indicates which signals induce $H$:
\begin{table}[ht]
\label{tab:types}
\begin{center}
\begin{tabular}{lccc}
\toprule
Type & Signals inducing $H$ & $\PP_H(H\text{ action})$ & $\PP_L(H\text{ action})$\\\midrule
A & Strong $H$ only & $9/14$ & $1/14$\\
B & Moderate and strong $H$ & $6/7$ & $1/7$\\
C & All except strong $L$ & $13/14$ & $5/14$\\\bottomrule
\end{tabular}
\caption{Types of signals and what they induce}
\end{center}
\end{table}

The associated action likelihood ratios for $(L,H)$ are $(5/13,9)$ in type A, $(1/6,6)$ in type B, and $(1/9,13/5)$ in type C. Multiplication by starting odds gives every transition in \Cref{tab:transitions}.

\begin{table}[ht]
\centering
\label{tab:transitions}
\begin{tabular}{rcrr}\toprule
Starting odds & Type & Odds after $L$ & Odds after $H$\\\midrule
$1$ & A & $5/13$ & $9$\\
$5/13$ & A & $25/169^{\ast}$ & $45/13$\\
$9$ & B & $3/2$ & $54^{\ast}$\\
$45/13$ & B & $15/26$ & $270/13$\\
$3/2$ & B & $1/4^{\ast}$ & $9$\\
$15/26$ & A & $75/338^{\ast}$ & $135/26$\\
$270/13$ & C & $30/13$ & $54^{\ast}$\\
$135/26$ & B & $45/52$ & $405/13^{\ast}$\\
$30/13$ & B & $5/13$ & $180/13$\\
$45/52$ & A & $225/676^{\ast}$ & $405/52$\\
$180/13$ & C & $20/13$ & $36^{\ast}$\\
$405/52$ & B & $135/104$ & $1215/26^{\ast}$\\
$20/13$ & B & $10/39^{\ast}$ & $120/13$\\
$135/104$ & B & $45/208^{\ast}$ & $405/52$\\
$120/13$ & B & $20/13$ & $720/13^{\ast}$\\\bottomrule
\end{tabular}
\caption{All higher-payoff informative odds. A star marks an absorbing destination.}
\end{table}

All unstarred states lie strictly between $k_2/9=10/27$ and $9k_2=30$. Every starred destination below one is strictly below $10/27$, and every starred destination above one is strictly above $30$. Thus the table is closed, includes all reachable informative states, and has no on-path ties. From every informative state, at most three consecutive $L$ actions reach an $L$ cascade. Each such action has probability at least $1/14$ under either state, so absorption occurs almost surely.

The following elimination reproduces the exact probabilities without an infinite-history calculation. Fix a true state and denote the three type-specific probabilities of $H$ by $a,b,c$, respectively. Let $f(o)$ be the probability of an eventual $L$ herd starting at odds $o$. In an $L$ cascade set $f=1$; in an $H$ cascade set $f=0$. Each table row gives the equation
\[
f(o)=(1-p)f(o_L)+pf(o_H),
\]
where $p$ is $a,b,$ or $c$ according to its type. Define
\begin{align*}
D&=1-b(1-b),&
r&=\frac{(1-b)^2}{D},\\
A&=1-a+a(1-b)(1-a+ar),&
B&=\frac{(1-c)(1-b)}D,\\
U&=\frac{1-a+a(1-b)A+ab^2(1-c)B}
           {1-ab(1-c)(1-b)}.
\end{align*}
The table equations give $r=f(9)=f(405/52)$, $A=f(15/26)$, $B=f(180/13)$, and $U=f(5/13)$. Consequently,
\begin{equation}\label{eq:renewal}
f(1)=(1-a)U+ar.
\end{equation}
Substituting $(a,b,c)=(9/14,6/7,13/14)$ under $H$, and $(1/14,1/7,5/14)$ under $L$, yields
\begin{equation}\label{eq:newprob}
\PP_H(\mathcal L_{v_2})=\frac{25669}{164260},\qquad
\PP_L(\mathcal L_{v_2})=\frac{32327}{32852}
=\frac{161635}{164260}.
\end{equation}
The difference $e_H(v_1)-e_H(v_2)=1071/164260$ is strictly positive. At the equal prior, Bayes' rule gives
\[
C_L(v_2)=\frac{161635}{161635+25669}
=\frac{161635}{187304}>\frac67=C_L(v_1),
\]
with difference $7621/1311128$. Finally, $e_L(v_2)=525/32852<1/43$, with difference $1195/164260$.

\subsubsection{Blackwell comparison}\label[appendix]{app:Blackwell}
The state-conditional probabilities of an $H$ herd are
\[
\begin{array}{c|cc}
&\PP_H(\mathcal H_v)&\PP_L(\mathcal H_v)\\\hline
v_1&36/43&1/43\\
v_2&138591/164260&2625/164260.
\end{array}
\]
Given the higher-payoff herd report, produce a lower-payoff $H$ report with probability
\[
\alpha=\frac{5793191}{5846538}\quad\text{after }H,
\qquad
\beta=\frac{14697}{1948846}\quad\text{after }L.
\]
Both probabilities lie strictly between zero and one. Direct substitution verifies
\begin{align*}
\alpha\frac{138591}{164260}+\beta\frac{25669}{164260}&=\frac{36}{43},\\
\alpha\frac{2625}{164260}+\beta\frac{161635}{164260}&=\frac1{43}.
\end{align*}
This state-independent garbling establishes dominance of the herd experiment.

To obtain the lower-payoff full record, next draw a lower-payoff record conditional on the garbled herd direction. This conditional law is independent of the state: all lower-payoff records ending in $L$ have terminal odds $1/6$, and all those ending in $H$ have odds $36$. Formally, under the equal prior the terminal posterior is constant on each herd event, so Bayesian calibration implies conditional independence of the state and full record given the herd. We may therefore use its conditional record law under the common prior. The resulting simulation reproduces the lower-payoff public-history laws under both states.

The comparison is strict. The posterior conditional on a higher-payoff $H$ herd is
\[
\frac{138591}{138591+2625}=\frac{46197}{47072}>\frac{36}{37}.
\]
Every lower-payoff full-record posterior is at most $36/37$, and a garbling can only take conditional averages of these posteriors. Thus it cannot reproduce the higher-payoff herd experiment or full-record experiment. This proves both strict Blackwell comparisons in \Cref{prop:confidence}.

By the prior-shift equivalence in \Cref{sec:comparisons}, the same example also
gives a reversal in the prior with zero expressive payoffs. Increasing the prior
probability of $H$ from $3/13$ to $2/7$ raises the probability of an eventual
$L$ herd conditional on $H$ from $25669/164260$ to $7/43$.

\subsection{Intervals of payoff comparisons}
\label[appendix]{app:reversalintervals}

The public-history laws remain unchanged over open intervals
of expressive payoffs. To establish this, it suffices to verify
that each private signal induces the same action at every
reachable public odds value, including absorbing states.

At a reachable odds value $o$, let $\ell(o)$ be the largest
private likelihood ratio assigned to $L$ in the baseline
regime, and let $u(o)$ be the smallest assigned to $H$.
Use $\ell(o)=0$ when all signals induce $H$, and
$u(o)=\infty$ when all signals induce $L$. The same signal
classification is preserved without ties whenever
\[
o\ell(o)<k<ou(o).
\]

For the lower-payoff regime, intersecting these restrictions
over the informative states $\{1,6\}$ and absorbing states
$\{1/6,36\}$ gives
\[
2<k<3.
\]
The binding lower restriction comes from the moderate $L$
signal at odds $6$; the binding upper restriction comes
from the moderate $H$ signal at odds $1$.

For the higher-payoff regime, applying the same calculation
to all informative states and absorbing destinations in
Table~\ref{tab:transitions} gives
\[
\frac{40}{13}<k<\frac{45}{13}.
\]
The binding lower restriction comes from the moderate $L$
signal at odds $120/13$, since
$(120/13)(1/3)=40/13$. The upper restriction comes from
the strong $H$ signal at odds $5/13$, since
$(5/13)9=45/13$. All remaining restrictions are weakly
less demanding.

Within either interval, each action therefore pools the same
signals at every reachable state. Starting from the same prior,
induction on the public history shows that action probabilities
and updated odds are unchanged. Hence the entire
state-conditional public-history laws are unchanged.

Finally, $v=(k-1)/(k+1)$ converts the two threshold intervals
into
\[
v_1\in(1/3,1/2),
\qquad
v_2\in(27/53,16/29).
\]
Every pair drawn from these intervals consequently satisfies
the same strict error, reliability, and Blackwell comparisons
as the baseline pair.

These intervals concern the discrete experiment. For each
fixed pair in their interiors, the atomless construction below also
preserves the corresponding public-history laws when its
bandwidth is sufficiently small.

\subsection{An atomless implementation}\label[appendix]{app:atomless}
The reference private posteriors associated with the four discrete signals are
\[
p_1=\frac1{10},\quad p_2=\frac14,\quad
p_3=\frac34,\quad p_4=\frac9{10}.
\]
Their equal-prior masses are $w_1=w_4=5/14$ and $w_2=w_3=1/7$. Replace each point $p_i$ by a uniform interval $[p_i-\varepsilon,p_i+\varepsilon]$ of the same mass. The unconditional density is
\begin{equation}\label{eq:atomless}
h_\varepsilon(p)=\sum_{i=1}^4\frac{w_i}{2\varepsilon}
\ind\{|p-p_i|\le\varepsilon\}.
\end{equation}
Set $f_H(p)=2ph_\varepsilon(p)$ and $f_L(p)=2(1-p)h_\varepsilon(p)$. These are probability densities because $h_\varepsilon$ is symmetric. Each interval has mean $p_i$ and mass $w_i$, so its conditional probabilities are $2w_ip_i$ under $H$ and $2w_i(1-p_i)$ under $L$. They exactly reproduce \Cref{tab:signals}.

It remains to verify that the intervals remain intact action categories. At odds $o$ and threshold $k$, the cutoff in reference posterior units is $k/(k+o)$. For the four lower-payoff states $\{1,6,1/6,36\}$, the smallest distance between any cutoff and any $p_i$ is $27/770$. For every higher-payoff informative state and absorbing destination in \Cref{tab:transitions}, the smallest distance is $9/2690$, attained at $o=405/13$, where the cutoff is $26/269$ and the nearest signal point is $1/10$. These claims are finite checks using the listed states.

Therefore, for every
\[
0<\varepsilon<\frac9{2690},
\]
all cutoffs remain outside the four signal intervals. Every action pools exactly the same categories as in the discrete construction, including at all absorbing states. Induction on the public history gives exactly the same state-conditional public-record laws, not just approximate error probabilities. One concrete choice is $\varepsilon=1/1000$.

The experiment is symmetric and atomless with actual likelihood-ratio endpoints $1/R_\varepsilon,R_\varepsilon$, where
\[
R_\varepsilon=\frac{9/10+\varepsilon}{1/10-\varepsilon}.
\]
All calculations in \eqref{eq:outcomes}, both strict Blackwell comparisons, and both primary-outcome reversals therefore hold for this atomless experiment. Its density is concentrated in separated evidence categories, so it lies outside the single-peaked benchmark.

\section{Overshooting and sufficiently separated payoffs}\label[appendix]{app:overshoot}
\subsection{The uniform bound}
Define
\begin{equation}\label{eq:constant}
m(c)=\EE_L[\lambda\mid\lambda\ge c],\qquad
C=\max_{c\in[1/R,R]}\frac{m(c)}c,\qquad \kappa=C/R,
\end{equation}
where $m(R)=R$ by continuous extension. The state-$L$ expectation is the action likelihood ratio, as in \eqref{eq:updateH}. At normalized starting odds $1/c$, the odds following $H$ are $m(c)/c$. Thus $C$ is the supremum of normalized odds produced by an informative $H$ action, and $\kappa$ measures the largest proportional overshoot of the boundary $R$.

\Needspace{17\baselineskip}
\begin{lemma}\label[lemma]{lem:overshoot}
For every symmetric atomless bounded experiment,
\begin{equation}\label{eq:Crange}
R\le C<R^2,\qquad1\le\kappa<R.
\end{equation}
The no-overshooting condition holds if and only if $C=R$. In particular, no delayed cascades implies $C=R$. Under \eqref{eq:active},
\begin{align}
k/C\le O_\infty\le k/R&\quad\text{on }\mathcal L_v,\label{eq:Lband}\\
kR\le O_\infty\le kC&\quad\text{on }\mathcal H_v.\label{eq:Hband}
\end{align}
\end{lemma}
\begin{proof}
At $c=1/R$, the upper pool includes all signals and $m(c)=\EE_L\lambda=1$, so $m(c)/c=R$. For $c>1/R$,
\[
\frac{m(c)}c\le\frac Rc<R^2.
\]
Atomlessness makes the numerator and denominator defining $m(c)$ continuous for $c<R$, even across gaps in the support. The denominator is positive because $R$ is the actual essential upper endpoint. As $c\uparrow R$, the conditional mean lies between $c$ and $R$ and converges to $R$. Thus $m(c)/c$, with its stated extension at $R$, is continuous on the compact interval $[1/R,R]$. Its maximum is attained, is at least $R$, and is strictly smaller than $R^2$. This proves \eqref{eq:Crange}.

For any informative history, \eqref{eq:normalized} implies $O_H^+/k\le C$. By reflection, $O_L^+/k\ge1/C$. An $H$ action increases odds and cannot first enter an $L$ cascade, and an $L$ action decreases odds and cannot first enter an $H$ cascade. The first entry into either cascade therefore satisfies the corresponding band in \eqref{eq:Lband}--\eqref{eq:Hband}. If there is no finite entry, \Cref{lem:convergence} places the limit on the appropriate boundary, which also satisfies the band.

Finally, no overshooting is equivalent to $\Phi_H(y)\le R$ throughout the informative interval together with its reflected lower bound for $L$. Since $\Phi_H(1/c)=m(c)/c$ and the endpoint value already equals $R$, these inequalities hold exactly when $C=R$. No delayed cascades implies these weak inequalities.

\end{proof}

\subsection{Proof of the payoff comparison}

\begin{proof}[Proof of \Cref{thm:separation}]
Write $k_i=k(v_i)$ and $C=\kappa R$. By
\Cref{lem:overshoot}, terminal odds in regime $i$ lie in
$[k_i/C,k_i/R]$ conditional on an $L$ herd and in
$[k_iR,k_iC]$ conditional on an $H$ herd. The sufficient
condition $k_2/k_1\ge\kappa=C/R$ gives
\[
k_2/C\ge k_1/R,\qquad k_2R\ge k_1C.
\]
The first inequality separates the two $L$-herd odds supports in \eqref{eq:Lband}, and the second separates the two $H$-herd supports in \eqref{eq:Hband}. Since odds convert monotonically into beliefs, both conditional posterior supports are ordered. Thus $m_L(v_2)\ge m_L(v_1)$ and $m_H(v_2)\ge m_H(v_1)$. This implies the error comparisons, and $C_L=1-m_L$ proves the reliability comparison. Additional conditioning on the state cannot expand either support. If the sufficient condition is strict, the supports are separated by a positive gap, both mean comparisons are strict, and all displayed outcome comparisons are strict.
\end{proof}
\begin{proof}[Proof of \Cref{cor:zero}]
The stated interval is equivalent to $\kappa<k(v)<R$. Since $\kappa<R$, it is nonempty. At prior $1/2$, both this regime and $k(0)=1$ are initially informative. We can then apply \Cref{thm:separation}.
\end{proof}

\subsection{Overshooting in the four-signal example}

For the discrete experiment, define the overshoot factor using
a supremum:
\[
\kappa=\frac1R\sup_{c\in[1/R,R]}
\frac{\EE_L[\lambda\mid\lambda\ge c]}{c}.
\]
The supremum accommodates jumps when the cutoff passes an atom.
Here $R=9$, and the conditional means are
\[
m(c)=
\begin{cases}
1, & c=1/9,\\
13/5, & 1/9<c\le1/3,\\
6, & 1/3<c\le3,\\
9, & 3<c\le9.
\end{cases}
\]
Consequently,
\[
\sup_t\frac{m(t)}t
=\max\left\{9,\frac{117}{5},18,3\right\}
=\frac{117}{5},
\qquad
\kappa=\frac{13}{5}.
\]

To verify the corresponding lower bound, the mean likelihood
ratios in the $L$ pool are $1/9$, $1/6$, and $5/13$ on the
three respective cutoff intervals above. Since starting odds
are $k/c$, every informative update satisfies
\[
O_H^+\le\frac{117k}{5},
\qquad
O_L^+\ge\frac{5k}{117}.
\]
An $L$ action cannot enter an $H$ cascade, and an $H$ action
cannot enter an $L$ cascade. Thus terminal odds satisfy
\[
\frac{5k}{117}\le O_\infty\le\frac{k}{9}
\quad\text{on }\mathcal L_v,
\qquad
9k\le O_\infty\le\frac{117k}{5}
\quad\text{on }\mathcal H_v.
\]
Absorption occurs almost surely: from any informative belief,
five consecutive $H$ actions suffice to reach an $H$ cascade,
and this sequence has probability at least $(1/14)^5$ under
either state.

For $v_1=3/7$ and $v_3=3/4$, we have $k_1=5/2$ and $k_3=7$.
The terminal-odds ranges are strictly ordered because
\[
\frac{5k_3}{117}=\frac{35}{117}>
\frac{5}{18}=\frac{k_1}{9},
\qquad
9k_3=63>\frac{117}{2}=\frac{117k_1}{5}.
\]
Hence both herd-conditioned mean beliefs in $H$ strictly
increase. The same accounting identities used in the proof
of \Cref{thm:separation} yield
\[
e_H(3/4)>e_H(3/7),
\qquad
C_L(3/4)<C_L(3/7).
\]

\end{document}